\documentclass[aps, amsmath,amssymb, amsfonts, nofootinbib, notitlepage, twocolumn, preprintnumbers]{revtex4-1}
\usepackage{graphicx}
\usepackage{bm}
\usepackage{bbold}

\graphicspath{ {./Desktop/} }

\begin{document}

\title{On the nature of an emergent symmetry in QCD with low-lying Dirac modes removed}

\author{Thomas D. Cohen}
 \email{cohen@physics.umd.edu}
 \affiliation{%
Department of Physics and Maryland Center for Fundamental Physics\\University of Maryland, College Park, Maryland\\
}%

\date{\today}

\begin{abstract}
Remarkable symmetry properties appear to arise in lattice calculations of correlation functions in which the lowest-lying eigenmodes of the Dirac operator in quark propagators are removed by hand. The Banks-Casher relation ties the chiral condensate to the density of low lying modes;  thus, it is plausible that  removal of such modes could lead to a regime where spontaneous chiral symmetry breaking does not occur.    Surprising, a pattern of identical correlation functions  was observed that is larger  than can be explained by a restoration of chiral symmetry.  This suggests  that a larger symmetry---one that is not present in the QCD lagrangian---emerges when these modes are removed.  Previously it was argued that this emergent symmetry was SU(4).   However,  when the low-lying modes are removed, the correlation functions of  sources in the  SU(4) 15-plet of  spin-1 mesons  appear to coincide with the correlation function of the SU(4) singlet.    A natural explanation for this is an emergent symmetry larger than SU(4).   In this work, it is  shown that there exists no continuous symmetry whose generators in the field theory are spatial integrals of local  operators that can account for the full pattern of  identical correlation functions unless the apparent coincidence of the singlet channel with the 15-plet is accidental.    \end{abstract}

\maketitle
\newcommand{\updownarrows}{\mathbin\uparrow\hspace{-.5em}\downarrow}
\newcommand{\downuparrows}{\mathbin\downarrow\hspace{-.5em}\uparrow}

\section{Introduction}

It has long been accepted that quantum chromodynamics (QCD) is the theory underlying strong interaction physics. However, due to its intrinsically nonperturbative character many of its fundamental features remain elusive.  Two aspects of QCD dynamics, confinement and spontaneous chiral symmetry breaking are critical to our understanding of the theory.   However, the relationship between the two has been murky.  Recent lattice gauge calculations\cite{DGL1,DGL2,GLS,DGP1,G1,DGP2,GP1} may  have shed significant light on this relationship.  These calculations are quite remarkable in that they suggest that in a regime where effects of chiral dynamics are turned off by hand (and thus confinement dynamics may be expected to dominate) a new symmetry emerges which is larger than any  explicitly in the QCD lagrangian\cite{G1,GP1,DGL2,DGL1,DGP1,DGP2}.   The nature of this putative emergent symmetry is the subject of the present paper.  For simplicity of analysis and to match onto existing lattice calculations, this paper concentrates on the case where there are two degenerate light flavors.

The proposal that a new symmetry emerges from QCD is quite  radical.  It was not introduced lightly\cite{G1,DGL2,DGL1,DGP1,DGP2}.  Rather, it was suggested to explain a totally unexpected phenomenon recently observed in lattice calculations of correlation functions with various sources including those with the quantum numbers of spin zero, spin one and spin two mesons and of baryons.  These correlation functions were computed only after the contributions from some number of modes with the smallest eigenvalues of the Dirac operator were removed.  The Banks-Casher relation\cite{BC} implies that the chiral condensate---an order parameter for chiral symmetry breaking---must vanish when the low-lying modes are removed.  This suggests that the removal of such modes might yield a regime which is chirally restored;  {\it i.e.} one in which correlation functions of sources connected  to each other by chiral transformations are identical.   Moreover, for most channels  a plateau in the effective mass plot continues to exist after the removal of these modes \cite{DGL1,DGL2,GLS,DGP1,G1,DGP2} which may be interpreted as suggesting that reasonably well defined hadronic resonances survive the truncation.   However, what was found when the correlation functions were calculated was totally unexpected: a set of identical correlation functions that was much larger than could be explained by chiral symmetry alone.  The proposal of a new emergent symmetry in QCD, was made to explain this remarkable fact.  
  
The emergent symmetry proposed to explain this pattern was SU(4); it was interpreted as a symmetry of the confining part of the QCD\cite{G1,DGP1,DGP2,GP1}.  It was suggested that an SU(4)  symmetry associated with the color-Coulomb interaction might ultimately be responsible for it\cite{G1,GP1}.  This paper takes an agnostic view of these interpretations.    However,  the paper takes seriously the possibility  that such an emergent symmetry exists in this regime, despite the radical nature of the proposal, given its significant explanatory power. 

However, there remains a puzzle associated with this scenario. The pattern of correlation functions that become identical upon removal of quasi-zero modes appears to be mostly explained by an emergent SU(4) symmetry.  However, it does not appear to explain the full pattern.  In particular, one of the J=1 correlation functions (an SU(4) singlet channel) was observed to be nearly identical to others (in an SU(4) 15-plet)\cite{DGL2} even though this is  ``accidental'' from the perspective of SU(4)\cite{GP1}.    In this paper a natural solution to this puzzle is explored: namely that the emergent symmetry group is  larger than SU(4) (and contains SU(4) as a subgroup).  However, as will be shown here, the nature of any such emergent symmetry must be rather peculiar: there is no symmetry transformation  of the usual kind for a continuous symmetry in a field theory (i.e. one whose  generators are given as the spatial integral of a local operator) that can explain the full  pattern of identical correlation functions (assuming that the singlet really is identical to the others).    In effect, what will be demonstrated here is a no-go theorem.  The theorem does not rule out an emergent symmetry that can explain the full pattern, but it does imply that such an emergent symmetry must be of a peculiar non-local sort.

One possibility, is that the correlation function in the singlet channel is, in fact, distinct from the others, but is accidentally so close that it is hard to distinguish the differences (given the quality of the numerics in the presently available lattice studies).  If this is true, then a remarkable coincidence has occurred.   If, however, future higher quality lattice studies continue to suggest that the singlet correlation functions and those 15-plet are the same, then the no-go theorem proved in this paper implies that the nature of the emergent symmetry is rather mysterious.  Thus, the no-go theorem for the usual type of symmetry transformation may hold important clues as to the  nature of the emergent symmetry.

This paper is organized as follows: in the next section, background is provided on the motivation for the proposed SU(4) symmetry of ref. \cite{G1,GP1,DGP1,DGP2}.  The emphasis of this section will be on channels with $J=1$ for which the most complete set of correlation functions have been computed.  The nature of the SU(4) symmetry   will be clarified in the following section.  Next is a section discussing the   puzzle and its possible solution due to an emergent symmetry larger than SU(4).  In the following section it is shown that there does not exist an emergent symmetry based on local currents that can account for the observed pattern unless the singlet channel is merely accidentally close to the others.   The paper concludes with a section which discusses some implications of this result and the outlook for future work.

\section{Background}

To begin let us review the lattice calculations and how they have been interpreted previously\cite{DGL1,DGL2,GLS,DGP1,G1,DGP2,GP1}.  The calculations in question are of correlation functions for gauge-invariant local  sources associated with various quantum numbers.  These are done in a standard manner with one important change---the contributions of some fixed number of eigenvectors of the Dirac operators are removed from quark propagators connected to the sources; the removed modes are those with the smallest eigenvalues.   These recent calculations were an out growth of the analysis done in ref.~\cite{LS}, which focused on chiral symmetry breaking effects. In these works, the quark propagator in any gauge configuration $A$,  $S_A$, was replaced  in these calculations by a modified propagator $S_A^M$:
\begin{equation}
S_A^M = S_A - \sum_{j=0,N}\frac{ |\psi_j \rangle \langle \psi_j|}{i \lambda + m} \; \; 
{\rm where} \; \; D\!\!\!\!/ \,  |\psi_j\rangle  = i \lambda_j |\psi_j\rangle
\end{equation}
and the sum is taken over the $N$  modes with the smallest values of $|\lambda|$.  

Of course, the effects of removing low-lying modes must depend on how many modes are are removed and, presumably, also on other parameters of the system such as the box size.  The interesting new effects observed in these lattice studies--a pattern of correlation functions in different channels becoming nearly identical---are seen to set in quite rapidly.  The tendency of correlators to become similar to one another is quite clear when as few as 10 modes are removed and the correlators have become nearly identical by the time 30 modes are removed. Physically for these lattices the removal of ten modes corresponds to removing modes with a virtuality of around 65 MeV or less  while the removal of 30 modes corresponds to removing modes with a virtuality of around 180 MeV or less.  As discussed in the concluding section, future lattice studies with a number of different volumes would be needed to pin down precisely the conditions for which interesting new qualitative effects emerge as modes are removed.

In these lattice studies, gauge configurations themselves were computed using an unmodified fermion determinant.   It is interesting to speculate that the removal of such modes in the propagator could have the effect of removing the effects of  spontaneous chiral symmetry breaking.  This speculation follows from the Banks-Casher relation\cite{BC} which is known to hold in the infinite volume limit for zero quark mass:
\begin{equation}
\langle \overline{q} q \rangle = - \pi  \langle \rho^D(0) \rangle  \; .
\label{Banks}
\end{equation}
In Eq.~(\ref{Banks}),  the chiral condensate,$\langle \overline{q} q \rangle$, is an order parameter for spontaneous chiral symmetry breaking. The density of states of the Dirac operator for gauge fields, $\rho^D(\lambda)$ is a continuous function at infinite volume and the brackets on the righthand side of Eq.~(\ref{Banks}) indicates averaging over gauge fields appropriately weighted in the functional integral.  At finite volume, the smooth density of states at zero  gets converted into a set of discrete levels near zero.  Their removal should, in effect, force the chiral condensate to vanish as one would expect in a chirally-restored regime.

It is noteworthy that these lattice studies treat ``valance'' and ``sea'' quarks on a different footing.  The quasi-zero modes are removed from the propagators connected to sources---the valance quarks---but not from the those in the fermion determinant associated with internal quark loops.  This, breaks various connections between observables. For example, the chiral condensate calculated via a propagator loop will vanish as $m_q \rightarrow 0^+$ when the low lying modes are removed.  However, the chiral condensate  can also be obtained from the
\begin{equation}
\langle \overline{q} q \rangle =  -\frac{1}{V} \frac{d \log \left (Z(m_q)\right )}{d m_q}
\end{equation}
where $V$ is the 4-volume and $Z(m_q)$ is the Euclidean space partition function and depends only on the treatment of the sea and, accordingly,  will not vanish in the chiral limit of $m_q \rightarrow 0^+$.  Thus, it is not totally clear {\it a priori} whether the removal of the low-lying modes in the propagators really does force the system into a chirally unbroken regime.

Consider the following two tests of the speculation that the removal of the low-lying modes creates a chirally-restored regime:  i)  In a chirally restored regime one expects that two-point correlation function for sources that are related by chiral transformations---for example, the $\pi$ and $f_0$ correlation functions sourced by $i \overline{q} \gamma_5 \vec{\tau} q$ and $i \overline{q}  q$ respectively---to be identical in a chirally restored phase. ii)  One also expects that  mixed correlation functions for two sources which share flavor, spin and parity quantum numbers but which correspond to different chiral representations to vanish in a chirally-restored regime while being generically nonzero if chiral symmetry is broken\cite{CJ}.  For example, the $\rho$ meson can couple to the source $\overline{q} \gamma_\mu \tau_a q$ which transforms in the chiral representation (0,1) + (1,0) or to the source $\overline{q} \gamma_\mu \gamma_0 \tau_a q$ which transforms in the chiral representation ($\frac{1}{2}$,$\frac{1}{2}$).  In a chirally broken regime, a mixed correlation function with these two sources is generically non-zero, while in a chirally restored regime it will vanish.  Both  effects i) and ii) are observed to be true to good accuracy in the calculations\cite{DGL1,DGL2}.  Thus, there are grounds for believing that the regime studied when the quasi-zero modes are excluded acts as though it is chirally restored.

The removal of the low-lying Dirac modes  yielded a pattern of correlation functions consistent with the  restoration of the SU(2)$_L \times$SU(2)$_R$ chiral symmetry.  One might expect a larger  pattern than this.   Apart from the SU(2)$_L \times$SU(2)$_R$ chiral symmetry,  the QCD Lagrangian, contains another chiral symmetry, associated with flavor-neutral axial rotations.  Of course, due to an anomaly, U(1)$_A$ is not a symmetry of the quantum theory.  The anomalous breaking of the U(1)$_A$ is of a fundamentally different character than the spontaneous breaking of the SU(2)$_A$ part of the the SU(2)$_L \times$SU(2)$_R$ chiral symmetry;  it acts as an explicit symmetry breaking and does not yield a Goldstone mode\cite{U1}.  However, it is well known that the U(1)$_A$  anomaly, like spontaneous breaking of SU(2)$_A$ is deeply tied to the physics of zero modes.  From the the anomaly, the space-time integral over the divergence of the axial current in the massless theory is proportional to the winding number $\nu$.  Moreover,  an index theorem\cite{U1}, ties the winding number to the  zero modes of the Dirac operator:
\begin{equation}
\nu = N_+-N_-
\end{equation} 
where $N_+$ and $N_-$ are the positive- and negative-chirality zero modes of the Dirac operator.  Thus, it might not be too  surprising if effects of the anomaly that act to split channels (that would otherwise be the same) might turn off in  calculations that remove explicitly the low-lying modes.   In fact, the lattice calculations of ref.~\cite{DGL1,DGL2,GLS,DGP1,DGP2}  are consist with this: correlation functions in channels connected to each other by U(1)$_A$ transformations, also appear to be nearly  identical when low-lying modes are removed.
\begin{center}
\begin{table}[b]
\begin{center}
	\begin{tabular}{|c|c|c|c|}
	\hline
channel & source & $I, J^{PC}$ & parity-chiral rep\\
     \hline
     \hline
  $f_1$ &$  i \overline{q} \gamma_j \gamma_5 q $&$0,1^{++}$& (0,0)\\ 
     \hline 
     \hline
      $ \omega $& $\overline{q} \gamma_j q $& $0,1^{--}$& (0,0)\\
     \hline
     \hline
      $\omega'$ &$ i \overline{q} \gamma_j\gamma_0 q $&$0,1^{--}$& ($\frac{1}{2},\frac{1}{2}$)\\
      $b_1$& $ \overline{q} \gamma_0 \gamma_5 \gamma_j \vec{\tau} q $&$0,1^{+-}$& \\
     \hline
      $\rho'$ &$i  \overline{q} \gamma_j \gamma_0 \vec{\tau} q $&$1,1^{--}$& ($\frac{1}{2},\frac{1}{2}$)\\
      $h_1$& $ \overline{q} \gamma_0 \gamma_5 \gamma_j q $&$0,1^{+-}$& \\
     \hline
     \hline
      $\rho$ &$ \overline{q} \gamma_j \vec{\tau} q $&$1,1^{--}$& (0,1)$\oplus$(1,0)\\
      $a_1$& $ i \overline{q} \gamma_j \gamma_5  \vec{\tau} q $&$1,1^{+-}$&  \\
     \hline
   \end{tabular}
\end{center}
\caption{Quark bilinear sources for various spin 1 channels with their parity-chiral representations.  $\omega'$ and $\rho'$ indicate operators with $\omega$ and $\rho$ quantum numbers but which are in distinct chiral multiplets from the standard $\rho$ and $\omega$ sources.   Operators that are not separated by a horizontal line ({\it eg.} $\rho$ and $a_1$) are connected to each other via SU(2)$_L \times$SU(2)$_R$  chiral transformations; their correlation functions should be identical in a chirally-restored regime.  Operators that are not separated by a double horizontal line ({\it eg.} $\omega',\, b_1, \,\rho',$ and $h_1$) are connected to one another via combination of SU(2)$_L \times$SU(2)$_R$  and U(1)$_A$ transformations; their correlation functions should be identical in a chirally-restored regime in which anomaly effects are turned off.  }
\label{table1} 
\end{table}
\end{center}

While this behavior is interesting, nothing about it so far is particularly mysterious.  The removal of the low-lying modes, yields a pattern  consistent with an unbroken SU(2)$_L \times$SU(2)$_R \times$U(1)$_A$ chiral symmetry in which the effects of both the anomaly and spontaneous symmetry breaking are turned off.  However, at this stage, a very surprising fact is encountered: the pattern of identical correlation functions seen in the lattice data of ref. \cite{DGL1,DGL2,GLS,DGP1,G1,DGP2,GP1} turns out to be much larger than this. 

 For example, consider the spin 1 mesons sourced by quark bilinears with no derivatives.  These sources are given explicitly in Table~\ref{table1}) along with their parity-chiral representations ({\it i.e.} chiral representations or sets of chiral representations that contain states of good parity).\footnote{ While several of these operators transform  under Lorentz transformations as the spatial components of 4-vectors, others (those containing a $\gamma_0$) are the one-spatial-one-temporal components of rank-2 Lorentz tensors.  It is not a problem that these operators correspond to different Lorentz structures.  Note, that if one constructs a state at rest by integrating the source over space at a fixed time slice and acting on the vacuum, all of these sources create a state with angular momentum of unity; that is all of these operators create vector or axial-vector mesons. }  
 
 If one assumes that correlation functions of sources connected by SU(2)$_L \times$SU(2)$_R \times$U(1)$_A$ transformations are identical, one would expect that that there would be four distinct classes of correlation functions for the sources in the list in Table~\ref{table1}) with identical correlation functions within each class but not between classes.  Class 1) contains the  $\omega$; class 2) contains the  $f_1$; class 3) contains a distinct $\omega$, the $b_1$, the $\rho$ and the $h_1$; class 4) contains another $\rho$ and the $a_1$.   However,  the correlation functions  in all four classes appear to be very nearly identical to each other once the modes with 30 lowest eigenvalues are removed\cite{DGL1,DGL2}.  The phenomenon of a larger class of identical correlation functions than  obtained from invariance under  SU(2)$_L \times$SU(2)$_R \times$U(1)$_A$  chiral transformations is also observed in tensor mesons\cite{DGP1}; however correlation functions for  the complete set of J=2 sources have not yet been calculated.   Similarly, a larger set of nearly identical correlation functions are found in isoscalar meson channels \cite{DGL2} and  baryon channels\cite{DGP2}.
  

What is the origin of this remarkable pattern of nearly identical correlation functions?  The most natural explanation is symmetry.  But this immediately raises a problem: the symmetries of the QCD Lagrangian are known and no symmetry explicitly present in the QCD Lagrangian yields such a pattern.   An emergent symmetry that is not explicitly present in the QCD Lagrangian that arises only upon the truncation of the low-lying modes would solve this problem. 

Emergent symmetries are rather common features of physical systems.  Two examples are quite well known in the context of QCD.  One of these is the SU($2 N_h$) heavy-quark symmetry spin-flavor symmetry (where $N_h$ is the number of  heavy flavors) that emerges in the heavy quark limit\cite{IW1,IW2,IW3}.  Note that while the QCD Lagrangian is not invariant under the switching charm quarks into bottom quarks since they have different masses, certain properties of states with one unit of charm or bottomness become insensitive to the mass and spin of the heavy quark in the limit where $m_H/\Lambda_{\rm QCD} \rightarrow \infty$.  In the heavy quark limit, the heavy quark carries essentially all of the momentum of the state.  Thus, in the state's rest frame, the heavy quark is essentially at rest and acts as a static color coulomb source for the light degrees of freedom in the problem.  One can then swap heavy quark flavors without affecting the dynamics.  Moreover, the spin of the heavy quark does not matter since it is only affected by color magnetic interactions which are down by a factor of $\Lambda_{\rm QCD}/m_H$.  Another example of an emergent symmetry is the SU($2 N_f$) spin-flavor symmetry for baryons in the large $N_c$ limit of QCD with $N_f$ degenerate light flavors\cite{GS1,GS2,DM1,DM2,DJM1,DJM2,TDC96}.  Again this symmetry is not present in the QCD Lagrangian  but is a property of baryon states at large $N_c$.
 
 In a manner analogous to those examples, removing the low-lying modes from the quark propagator might  lead to a theory with a higher level of symmetry than seen in the Lagrangian.  It is important to note that in this context ``theory'' does not mean a well-defined unitary local quantum field theory.  Rather,  a ``theory''  can be thought of as a procedure that produces Euclidean-space correlation functions for specified sources.   To see how an emergent symmetry of this sort might come about, first imagine that the full action of QCD can be broken into two parts: one with some higher symmetry than is explicitly in the QCD Lagrangian and a second piece that explicitly breaks this higher symmetry.  Secondly, assume, that for some presently unknown reason, the part of the theory which breaks this higher symmetry couples predominantly---or solely---to the low-lying Dirac modes.  If this scenario is correct, a higher symmetry (either exact or approximate) will emerge  when the low-lying modes are truncated.   It is important to stress that at this stage that this scenario, while consistent with the lattice data and capable of producing an emergent symmetry, is entirely speculative.  It is not known how to break up QCD into two parts such that one part has a symmetry not present in the original Lagrangian while the other part couples predominately to low-lying modes.

 It might be objected that such a scenario is, by its nature, quite speculative, and {\it a priori} might seem to be far fetched.  However, it is important to recall that the phenomenon of identical correlation functions in larger patterns than can be explained by the known symmetries of QCD is both  quite striking and totally unexpected.  Presumably its explanation will require something that was also unexpected {\it a priori}.  In any case, in the absence of a more plausible explanation, it is natural to explore and test this one.
 
Accepting the scenario of an emergent symmetry as worth exploring, the next step is to ask what type of emergent symmetry can explain the data.  The approach taken in refs.~\cite{G1,GP1} was group theoretical.  Whatever group was relevant had to be large enough to contain SU(2)$_L \times$SU(2)$_R \times$U(1)$_A$ as a subgroup.  Since the pattern of  which  correlation functions become identical when quasi-zero modes are removed included states with the chiral group in the (0,0), two distinct ($\frac{1}{2}$,$\frac{1}{2}$) representations and the (0,1)+(1,0) representation, one is minimally searching for a group with a  single representation that encompasses these.  It was found that the minimum group that did this was SU(4).   A  study of the behavior of the sources in Table \ref{table1} under SU(4) transformations immediately yields a  pattern of   identical corrleators   that is far bigger than expected from SU(2)$_L \times$SU(2)$_R \times$U(1)$_A$.
 
\section{Symmetry operators}
  
It is useful  in interpreting the underlying physics to formulate how this SU(4) symmetry works explicitly.  It is important to stress here that while the putative symmetry is very strange from the perspective of QCD, in one critical respect it is utterly conventional: its generators are represented in the quantum field theory as the spatial integrals of local operators composed of the fields.

To see how this works, it is useful to start by constructing a realization of the generators of SU(4) in terms of the following basic objects which one may take as acting on quark fields:   the three Pauli matrices in isospin space, $\gamma_5$ and $\gamma_0$.  In this realization, the 15 generators of SU(4) are given by
\begin{equation}
\hat{Q}_a = \frac{1 }{2}  \int {\rm d}^3 x \, \hat{q}^\dagger(\vec{x},t)  \, \lambda_a \, \hat{q}(\vec{x},t)
\label{gen}\end{equation}
where the $\lambda_a$ are defined in Table \ref{table2} and the hat indicates a quantum mechanical operator;  $\hat{q}(\vec{x},t)$ is the quantum mechanical field operator.  Note that the generators are defined as integrals over a fixed time slice. The standard fermionic anti-commutation relations ensure that that the commutation relations of the $\hat{Q}_j $ are the same as that of the $\lambda_j$:
\begin{equation}
\left [ \hat{Q}_a ,\hat{Q}_b \right ] =  \hat{q}^\dagger(\vec{x},t) \left [  \frac{\lambda_a}{2} , \frac{\lambda_b}{2}    \right ] \hat{q}(\vec{x},t)
\label{q}\end{equation}

\begin{center}
\begin{table}
\begin{center}
\scalebox{.93}{
	\begin{tabular}{|c||c|c|c|c|}
	\hline
~&${ 1}$ & $\tau_x$ & $\tau_y$ &$\tau_z$\\
     \hline
     \hline
  ${ 1}$ & ~  &$ \lambda_1=\tau_x $& $\lambda_2=\tau_y$&$\lambda_3=\tau_z$\\ 
     \hline 
  $ \gamma_0$ & ~$\lambda_4 =   \gamma_0$  &$ \lambda_5=\gamma_0 \tau_x $& $\lambda_6=\gamma_0 \tau_y$&$\lambda_7=\gamma_0\tau_z$\\ 
  \hline
    $ \gamma_5$ & ~$\lambda_8 =   \gamma_5$  &$ \lambda_9=\gamma_5 \tau_x $& $\lambda_{10}=\gamma_5 \tau_y$&$\lambda_{11}=\gamma_5\tau_z$\\ 
    \hline
    $ i \gamma_5 \gamma_0 $ & ~$\lambda_{12} =   i \gamma_5 \gamma_0$  &$ \lambda_{13}=  i \gamma_5 \gamma_0 \tau_x $& $\lambda_{14}=  i \gamma_5 \gamma_0 \tau_y$&$\lambda_{15}=  i \gamma_5 \gamma_0\tau_z$\\ 
     \hline
   \end{tabular}
   }
\end{center}

\normalsize
\caption{The operators $\lambda_j$ used in defining the generators of SU(4) transformation in Eq. (\ref{gen}).  The $\lambda_a$ act in isospin and Dirac space.  The columns of the table indicate the isospin structure while the rows indicate the Dirac structure. }
\label{table2} 
\end{table}
\end{center}

One can exponentiate these generators to construct  quantum mechanical operators that act as group elements when acting in the quantum field theory:
\begin{equation}
\hat{U}(\vec{\alpha}) = \exp \left  ( i \sum_{a=1}^{15}  \alpha_a \hat{Q}_a  \right )
\label{groupel}\end{equation}
If is easy to show from the fermionic anticommutation relations that action of the SU(4) group on a quark field  operator takes the form,
\begin{equation}
\hat{U}(\vec{\alpha}) \, \hat{q} \, (\vec{x},t) \hat{U}^\dagger (\vec{\alpha})  =\exp \left ( \sum_{a=1}^{15} \frac{i \, \alpha_j \lambda_j }{2} \right ) \hat{q}(\vec{x},t) \; .
\label{form1}\end{equation}
 
Consider the action of the transformation on local quark bilinear sources with meson quantum numbers of the form
 \begin{equation}
 \hat{J}=\hat{q}^\dagger \Gamma_J \hat{q}
 \end{equation}
 where $\Gamma_J$ is a matrix in Dirac-isospin space.  Under a transformation in Eq.~(\ref{groupel}) it transforms in the following way:  
  \begin{equation}
  \begin{split}
& \hat{U}(\vec{\alpha})\hat{J}\hat{U}^\dagger(\vec{\alpha})=\\
&q^\dagger \exp \left ( \sum_{a=1}^{15} \frac{-i \, \alpha_a\lambda_a }{2} \right )\Gamma_J \exp \left ( \sum_{b=1}^{15} \frac{i \, \alpha_b \lambda_b}{2} \right ) \hat{q} \; .
 \end{split} \end{equation}

It should be clear that the transformations generated by Eq.~(\ref{groupel}) act to mix various sources together.  For example, consider the symmetry operator which corresponds to transformation in the  $\lambda_5$ direction: 
$\hat{U}_5(\theta) \equiv  \exp \left  ( i \theta \hat{Q}_5  \right )$.  If this acts on the $\omega$ source ($\overline{q} \gamma_j q$) it will yields a mixture of the $\omega$ and the  $\rho'$ ($ i  \overline{q} \gamma_j \gamma_0 \vec{\tau} q $):
\begin{equation} 
\begin{split}
 \hat{U}_5^\dagger(\theta) \left (\overline{q} \gamma_j q \right )  \hat{U}_5(\theta) &= \\
\cos(\theta) \left (  \overline{q} \gamma_j q \right )  & + \sin(\theta) \left ( i  \overline{q} \gamma_j \gamma_0 \vec{\tau} q \right ) \, .
\end{split}
\end{equation}
The key thing to note here is that the group operation acting on a source from Table \ref{table1} turned it into  a superposition of sources from Table \ref{table1}; it did not admix any other sources.  This property is generic: any  SU(4) group operation on any of the sources in Table \ref{table1} yields a superposition of sources in Table \ref{table1}:
\begin{equation}
\hat{U}(\vec{\alpha})\hat{J}_a \hat{U}^\dagger(\vec{\alpha}) = \sum_b C_{ab}(\vec{\alpha}) \hat{J}_b
\end{equation}
with $J_a$ representing the 16 vector currents in Table \ref{table1}.
That is,  the set of vector sources in Table \ref{table1} is closed under SU(4); the  operators in Table \ref{table1} transform as a representation of SU(4).  

Given these symmetry operators, one can define precisely the sense in which the theory has an SU(4) emergent symmetry.  One starts by denoting the Euclidean space correlation function  for a source $J$ as $\Pi_J(\vec{x},\tau)$.  The correlation function with quasi-zero modes removed is then denoted $\tilde{\Pi}_J(\vec{x},\tau)$. A  theory is said to have an emergent SU(4) symmetry if for all local currents $J$, 
\begin{equation}
\begin{split}
 \tilde{\Pi}_{J'_{\vec{\alpha}}}(\vec{x},\tau)  &=  \tilde{\Pi}_J(\vec{x},\tau) \; \; \; {\rm where}\\
 J'_{\vec{\alpha}}(\vec{x},\tau) & \equiv \hat{U}(\vec{\alpha})\hat{J}_a \hat{U}^\dagger(\vec{\alpha}) 
\end{split}
\label{ES} \end{equation}
for arbitrary SU(4) transformation $\vec{\alpha}$.  The emergent  symmetry can be said to be approximate if the equality in Eq. (\ref{ES}) only holds approximately.

 \section{A puzzle}
 
 The SU(4) symmetry proposed in refs.~\cite{G1,GP1,DGP1,DGP2} implies that correlation functions (computed with quasi-zero modes removed) for sources in any irreducible  SU(4) multiplet will be identical since all of these sources transform into each other under SU(4) transformations.  That is, every member of the multiplet can be transformed into any other member of the multiplet by a transformation of the form in Eq.~(\ref{groupel}) using the generators defined in Eq.~(\ref{gen}) and Table \ref{table2}.   As noted previously, the currents in Table \ref{table1} transform as an SU(4) representation.   At first glance, this may appear to explain why the correlation functions for the spin 1 sources in  Table \ref{table1} are identical---or nearly so---when computed with low-lying modes truncated.    
 
 However this is not the case.   While these sources do transform as an SU(4) representation, they do not transform as an  irreducible one.  Irreducible representations of SU(4) include, among others, a singlet, a fundamental of dimension 4, and an adjoint representation of dimension 15. There are 16 J=1 meson  sources in Table \ref{table1} once isospin is taken into account.  These can be grouped into  a 15-plet (adjoint representation) comprising all of the sources except for the $f_1$, and a singlet containing the $ f_1$.  It is easy to verify that the $f_1$ source transforms into itself under all SU(4) transformations.

Thus,  the expectation based on SU(4) is that the correlation functions for all members of the 15-plet will be identical but will be distinct from the $f_1$ correlation function in the singlet. However, as seen in ref. \cite{DGL2} the correlations functions of the members of the 15-plet, do indeed become identical to good accuracy after  low-lying modes are removed.  However, the data in  ref. \cite{DGL2}  suggests that  the $f_1$, the  singlet, also has a correlation function which, given the accuracy of the calculation, appears to be the same as the other 15 once the quasi-zero modes are truncated.  This is quite puzzling if SU(4) symmetry is the principal reason for why the correlation functions are the same.
 
One possible explanation for this is some sort of accident.  {\it A priori} this seems unlikely.   ``Accidental'' degeneracies that one comes across in physics are typically not truly accidental, but are due to a non-obvious symmetry.  Thus for example, the degeneracy in the spectrum of a Coulomb potential between states with the same principal quantum number but different $L$, which is accidental from the perspective of the rotational symmetry of the problem, is actually, due to the fact that the theory has  a hidden SO(4) symmetry.  

Before dismissing entirely the possibility the singlet and 15-plet correlation functions are essentially the same by accident, it is worth noting the limited numerical accuracy of the lattice studies.  One might imagine that if both the statistical and systematic uncertainties implicit in these lattice calculations were greatly reduced that perhaps the correlation functions of the members of 15-plet of J=1 operators under SU(4) would become increasing close to one another while  the $f_1$ singlet might be seen to be clearly different; the $f_1$ channel might be merely ``very close'' to the 15-plet for accidental reasons but not truly the same.  This seems unlikely, but not, perhaps unthinkable.  In this context, it is worth noting that the quality of the $f_1$ correlation function is particularly limited.  If one looks at the lattice data in ref.~\cite{DGL2}, the direct computations of the correlation functions based on a single source with the appropriate quantum numbers (Figure 10), then the $f_1$ channel appears to be as close to the other channels as they are to each other over approximately four decades.  However, the errors in the $f_1$ channel become substantial (much larger than the spread between the channels) far earlier than in the other channels, making it hard to say whether it remains very close to the others over a fifth decade.  If one looks at effective mass plots, computed using multiple sources (Figure 12 of ref.~\cite{DGL2}), the limitations of of the numerics become quite clear:  the $f_1$ is the only channel for which a clear plateau of more than 3 time steps is not apparent due numerical noise.  That said, the effective mass extracted from this short plateau coincides, within rather modest errors, to the effective masses of the other channels (Figure 13) of ref.~\cite{DGL2}.

If the explanation is not an accidental degeneracy, then  some alternative explanation is required.  The most naturally one is invariance under a larger emergent symmetry that contains SU(4) as a subgroup.  Such a symmetry must contain the singlet and adjoint representations of SU(4) in a single representation, since this is needed to explain the degeneracy observed.  However, as shown in the next section, there is no  generalization of the emergent SU(4) symmetry that can explain the pattern that is based on generators of usual type---ones that act on the QCD quark and/or gluon fields and are generated by the spatial integral of local operators constructed from quark and gluon fields.  

\section{A no-go theorem}

In this section a no-go theorem is established.  It demonstrates that no emergent symmetry connecting the $f_1$ to the other $J=1$ channels can exist: if  a)  its generators are spatial integrals of local dimension-three operators composed of QCD field operators and b)  it does not connect spin zero and spin one source operators.  A stronger no-go theorem dropping the restriction of dimension-3 operators is also demonstrated.   It is important to stress that, like all no-go theorems, these are of limited applicability.  They do not rule  out the possibility of an emergent symmetry.  However, they imply that any emergent symmetry capable of explaining the full pattern of which correlation functions become identical when quasi-zero modes are removed must be of an unusual type. 

 First the weaker no-go theorem  is established.  It applies to a possible emergent symmetry whose generators are spatial integrals of local dimension-three operators composed of QCD field operators.\footnote{Here ``dimension''  refers to  engineering dimension.  With typical symmetries associated with conserved currents the anomalous dimensions are zero and the distinction between the scaling dimension and the engineering dimension is irrelevant.  Here, the distinction may matter as the currents in question are not all conserved in the full quantum theory.  However, it should be clear in the present context that the engineering dimension is what is relevant: it controls the behavior at short distances (i.e. lattice scales) and the symmetry pattern is evident in correlation functions (with low lying modes removed) whose short distance (lattice-scale) sources are conjectured to be related by a symmetry. }    Such generators are special: their transformations map local operators into other local operators of the same dimension.  This is a very natural property for a symmetry operator and, it is not surprising that the symmetries one encounters in 3+1 dimensional field theories are typically of this form. 

In particular it is shown that no symmetry of this type that can connect the $f_1$ channel to the other $J=1$ channels in Table \ref{table1} exists, unless it also also connects the $J=1$ sources in Table \ref{table1} to $J=0$ channels  in Table \ref{table3}.  This means that such an emergent symmetry must have $J=0$  correlation functions (computed with the quasi-zero modes removed) equal to the $J=1$ correlation functions (similarly computed).  Since this is not seen in the lattice data, one concludes that no emergent symmetry of this form can explain the full pattern of which correlation functions are the same.  Next it is shown that relaxing the restriction to dimension-three  operators does not alter the result: no symmetry whose generators are spatial integrals of local operators can explain the pattern. 

 Consider generators that are spatial integrals of local dimension-three operators.   The key point is that an operator that affects the sources in Table \ref{table1} must contain quark fields---all purely gluonic operators commute with the entire set---and the number of local gauge-invariant dimension-three operators in QCD containing quark-fields is quite limited.  All such operators are of the form $\hat{q}^\dagger(\vec{x},t)  \Gamma \hat{q}(\vec{x},t)$ where $\Gamma$ is a matrix in Dirac-isospin space.  Since, Dirac-isospin space is 8 dimensional, there are  64 linearly independent operators. 

One of these 64 operators is the baryon density operator, $q^\dagger q$, whose generator is the baryon number and commutes with all  sources with mesonic quantum numbers; it is clearly not involved in any symmetry that mixes mesonic sources.    Another 15 of these are associated with the SU(4) symmetry already considered where $\Gamma$ is one half times one of the 15 $\lambda$ matrices from Table \ref{table2}.   In addition to these there is another set of three generators given by
\begin{equation}
\begin{split}
&\hat{S}_j = \frac{1 }{2}  \int {\rm d}^3 x \, \hat{q}^\dagger(\vec{x},t)  \,  \gamma_5 \gamma_0 \gamma_j \, \hat{q}(\vec{x},t) \; \; \; {\rm for} \; \; \; j=1,2,3\\
&{\rm with} \; \; \; [S_j, S_k] = i \epsilon_{j k l} S_l  \; . 
\end{split} \end{equation} 
These form an SU(2) algebra and, acting on $J=1$ operators of the form $\hat{q}^\dagger(\vec{x},t)  \Gamma \hat{q}(\vec{x},t)$  simply  generate ordinary spatial rotations.  Thus all of the sources in Table \ref{table1}, are triplets under this SU(2).  This SU(2) does not act to mix different kinds of $J=1$ sources; it merely changes the spatial direction within each type.  None of the preceding operators connect the $f_1$ source to any of the other $J=1$ sources and thus cannot by themselves explain the pattern of  which  correlation functions become identical when quasi-zero modes are removed.

The remaining 45 operators are of the form 
\begin{equation}
\hat{Q}_{a, j} = \frac{1 }{2}  \int {\rm d}^3 x \, \hat{q}^\dagger(\vec{x},t)  \,  \lambda_a \gamma_5 \gamma_0 \gamma_j \,  \hat{q}(\vec{x},t)
\end{equation}
where $\lambda_a$ is a matrix from Table \ref{table2}.  However, all of these---and any linear combination of them---can be ruled out as being generators of an emergent symmetry.  The reason is that these putative symmetry generators act to mix spin 1 sources in Table \ref{table1} with an isovector spin-zero sources (the $\pi$, $\pi'$, $\delta$ and $\delta'$ entries in Table \ref{table3}).   This can be seen directly from commutation relations.  For example,
\begin{equation}
[\hat{Q}_{13, j}, \hat{\overline{q}} \gamma_j \hat{q}]= - i \hat{\overline{q}} \tau_x \hat{q}
\end{equation}
which shows that the generator $\hat{Q}_{13, j}$ mixes an $\omega$ source polarized in the $j$ direction with the isovector scalar $\delta$ source.  This can be seen to be generic in that every $\hat{Q}_{a, j}$ operator acts to mix some spin 1 sources from Table \ref{table1} with an isovector spin-zero sources.    However if  the  45 $\hat{Q}_{a, j} $ operators are removed from consideration,  all of the possibilities have been exhausted.  There exists no emergent symmetry generated by spatial integrals of local dimension three operators that forces the correlation function for $f_1$ channel to be identical to the other $15$ spin-one channels when computed with quasi-zero modes removed unless isovector spin-zero correlation functions are also idenitcal.  This is the minimum form of the no-go theorem.

This no-go theorem is important since one can rule out the possibility that correlation functions of the isovector spin-zero channels  are identical to the correlation functions spin-one channels when calculated with quasi-zero modes removed based on lattice phenomenology.  Correlation functions of isovector J=0 sources have been studied with low-lying modes removed\cite{DGL2}.    While  all of these were found to be essentially equal to each other and consistent with an SU(4) symmetry,  they differ qualitatively  from the correlation functions of  $J=1$ sources.  In particular, these correlation functions, unlike those for $J=1$, have no plateau in the effective mass plot.   Thus, the spin 1 and spin 0 correlation functions cannot be connected by an  operator that acts as a generator for  an emergent symmetry.   Combined with the the minimal no-go threorem, this means that no  emergent symmetry whose generators are spatial integral of dimension three operators can explain the observed pattern of identical correlation functions in the J=1 sector.

\begin{center}
\begin{table}[b]
\begin{center}
	\begin{tabular}{|c|c|c|c|}
	\hline
channel & source & $I, J^{PC}$ & parity-chiral rep\\
     \hline
         \hline
      $\sigma$ &$ i \overline{q} \gamma_0 q $&$0,0^{++}$& ($\frac{1}{2},\frac{1}{2}$)\\
      $\pi$& $ \overline{q}  i \gamma_5  \vec{\tau} q $&$1,0^{-+}$& \\
     \hline
      $\delta$ &$ \overline{q} \vec{\tau} q $&$1,0^{++}$& ($\frac{1}{2},\frac{1}{2}$)\\
      $\eta$& $ \overline{q}  i \gamma_5 q $&$0,0^{+-}$& \\
     \hline
     \hline
       $\pi'$ &$ \overline{q}  \gamma_0 \gamma_5 \vec{\tau} q $&$1,0^{-+}$& (0,1)$\oplus$(1,0)\\
      $\delta'$& $  \overline{q} \gamma_0  \vec{\tau} q $&$1,0^{++}$&  \\
     \hline 
     \hline
  $\eta'$ &$ \overline{q} \gamma_0 \gamma_5 q $&$0,0^{-+}$& (0,0)\\ 
     \hline
     \hline
       $ \sigma' $& $\overline{q} \gamma_0 q $& $0,0^{++}$& (0,0)\\
     \hline
   \end{tabular}
\end{center}
\caption{Quark bilinear sources for various spin 0 channels with their parity-chiral representations. The $\sigma'$, $\pi'$, $\delta'$ and $\eta'$ indicate operators with the same quantum numbers  as $\sigma$, $\pi$, $\delta$ and $\eta$ respectively but which are in distinct chiral multiplets from the standard  sources.}
\label{table3} 
\end{table}
\end{center}

Next consider whether the  conclusion would be altered if the restriction that the local operators must be dimension three is dropped.   It is easy to show that  it is not.  

Since there are no local gauge invariant operators  in QCD with dimension less than three, one only need consider operators with dimension greater than three times constants with inverse powers of mass.  Thus the most general form
 of generator one needs to consider is
\begin{equation}
G = \int d^3 x \sum_{n=0}^\infty \frac{\rho^{3+n}(\vec{x})}{\Lambda^n} = \sum_{n=0}^\infty \frac{G^{(n)}}{\Lambda^n}
\label{G} \end{equation}
where $\rho^k$ is a local gauge-invariant operator of dimension $k$ and $\Lambda$ is a parameter with dimensions of mass\footnote{One might ask whether a more general form could be considered---namely ones in which a higher dimensional local operator has its dimension fixed by contractions with  appropriate powers of the position $\vec{x}$ as opposed to inverse powers of the mass.  An example might be $\vec{x} \cdot q^\dagger \vec{D} q$ where $D$ is a covariant derivative.  Operators such as these  have not been included in the analysis in this paper, since they seem unnatural: the  generators obtained by spatial integrals over these would depend on the choice of origin.  However, it is straightforward, if slightly more involved, to generalize the analysis to include them.  It is easy to see that the conclusions are not altered if one does this.}; some of the $\rho^k$ may may be zero.  Note that the individual terms ${G^{(n)}}/{\Lambda^n}$ are not necessarily symmetry generators on their own; the symmetry may require particular linear combinations of these.  Let us now consider, $J^D$, a source of dimension $D$, and how it transforms under unitary transformations generated by $G$.  It is easy to show in general that
\begin{align}
e^{i G \theta} J^D e^{-i G \theta} = & \sum_{n=0}^\infty \frac{J'^{D+n}}{\Lambda^n} \nonumber\\
{\rm with} \; \; \; J'^D=& e^{i G^{(0)} \theta} J^D e^{-i G^{(0)} \theta}  \label{G2}
\end{align}
where $ J'^D$ is a source with dimension $D$.   The key thing to note is that $J'^D$, depends only on the original sources, $\theta$ and $G^{(0)}$---the piece of the generator that is the spatial integral of a local dimension three operator.   This should be obvious:  higher $G^{(n)}$ always come with inverse powers of $\Lambda$ and to the extent they alter the source it will always lead to sources of higher dimensions multiplied by inverse powers of $\Lambda$.

A symmetry  explanation of why  the $f_1$ correlation function  should become identical with those of the other $J=1$ channels requires a transformation that can transform the $f_1$ sources into any of the other sources in Table \ref{table1}.   Equation (\ref{G2}) implies that two conditions for such a unitary transformation:  i)  all of the higher operators that could in principle be generated by the transformation ({\it i.e.} all terms  in the sum with $n \ge1$) must be zero and ii) for the generator of the transformation, $G^{(0)}$--the part which is the integral of a dimension-three operator--acting as a transformation on its own must  convert the $f_1$ into another $J=1$ sources from Table \ref{table1}.  It would be difficult to find a non-trivial transformation for which condition i) holds.  However, even if one could find  a transformation satisfying condition i), the no-go theorem obtained earlier applies to the transformation based on  $G^{(0)}$ in condition ii) as well.  The only transformations that can do this, necessarily also mix $J=1$ and $J=0$ sources and as such  are ruled out as viable candidates for an emergent symmetry.

\section{Discussion and outlook \label{disc}}

To summarize the puzzling situation:  The proposed emergent SU(4) symmetry of ref.\cite{G1,DGL2,DGL1,DGP1,DGP2} does not explain why the correlation function in the $f_1$ channel appears to be the same as the other $J=1$ channels when computed with quasi-zero modes removed.  Moreover, there is a no-go theorem that shows that any emergent symmetry based on a larger continuous group than SU(4) with generators that are spatial integrals of local operators, must mix J=0 and J=1 sources in a phenomenologically unacceptable way.   Ultimately, this means that to explain why the correlation function in the $f_1$ channel coincides with the other $J=1$ channels, requires the emergent symmetry not merely to have a different symmetry group than SU(4),  but to be a very different kind of symmetry.  

To put this puzzle in context it is important to recall that the observation of nearly identical correlation functions---when calculated with low-lying modes removed---in channels that are not connected by chiral symmetry is truly remarkable.  There is a real prospect that studies of such apparently identical correlation functions can give us a new tool to probe the nature of QCD dynamics and to gain important new insights into the theory.   For example, it has been suggested that the correlation functions computed with the quasi-zero modes removed give insights into the symmetries of the confining part of QCD and that the SU(4) of the color Coulomb interaction plays a key role\cite{G1,GP1}. However, before one can reliably draw inferences about QCD from such studies, the nature of this phenomenon needs to be understood better.

Clearly, sorting out what is happening will be greatly aided by more and better lattice studies.  It is important to verify the extent to which the correlation functions really do become identical when quasi-zero modes are removed and the precise conditions need for the phenomenon to happen.   High accuracy studies of the $J=1$ channels are need to ascertain whether or not the $f_1$ is  merely accidentally close to the other other $J=1$ channel correlation functions but is qualitatively different.   However, if  this turns out not to be the case, the situation is quite puzzling.

A key issue that needs clarification is the underlying cause for the pattern of  identical correlation functions.  Apart from the possibility of an emergent symmetry, there is no  known plausible explanation for the pattern.   Thus, it is sensible to take seriously the possibility that removing the low-lying modes exposes an emergent symmetry not immediately apparent in the underlying theory.  However there are two central problems that need to be addressed.  One is the principal subject of this paper, namely that it is not known to what group this emergent symmetry is associated---assuming that the $f_1$ channel really does become identical to the others.  Since any such emergent symmetry  will of necessity be nonstandard, a search for it could be particularly difficult.  A second issue  is that  that at present, there is no known reason why a symmetry that is not explicitly in the QCD Lagrangian should emerge at all.   This is in sharp contrast to other examples of emergent symmetries in QCD such as the emergent heavy-quark symmetry for situations when there is a a single heavy quark in the system or the emergent spin-flavor symmetry for baryons at large $N_c$.  If it is  correct that the degeneracies are caused by an emergent symmetry,  it is important to understand both which symmetry  emerges and why it does so.  Clearly, these two issues are related: if one can understand why an emergent symmetry exists one may well have clues as to what that symmetry might be and conversely, if one can identify the symmetry one may garner insights into what causes it to emerge.

Assuming that the conjecture of an emergent symmetry is correct and the SU(4)  singlet J=1  channel is identical to the other J=1 channels, it is clearly a daunting task to identify the emergent symmetry.   However, there is one obvious direction to pursue.  Unless spin 0 and spin 1 correlation functions becomes identical when quasi-zero modes are removed,   the no-go theorem excludes symmetries whose generators are spatial integrals of {\it local} operators.  Thus generators of  an emergent symmetry should contain {\it nonlocal} operators.   {\it A priori},  it is not implausible  that non-locaility is important given that  the operation of excluding quasi-zero modes from propagators is intrinsically  nonlocal when viewed in configuration space\cite{Nonlocal}.   That said, it is by no means apparent how to productively search for an emergent symmetry containing such nonlocal generators.

Of course, one perspective is that a symmetry as unusual as needed here is unlikely to be correct.  Given such a perspective, one could make a prediction: more accurate lattice studies of the $J=1$ channels with the lowest modes removed will show that the $f_1$ channel really behaves differently from the others.  The trouble with such a perspective however, is that the $f_1$ really is seen to behave like the others up to the quality of the numerics and it would be a very large coincidence for this to happen accidentally.  In any case,  this issue can be best resolved by doing more accurate lattice calculations.   Another, useful activity is to do lattice studies in other channels to see if other examples of correlation functions corresponding to distinct SU(4) multiplets appear to have identical correlation functions when computed with the low-lying modes removed.

There are other good reasons to put in the substantial effort that such lattice studies would involve:  there are some deep issues associated with the the possibility of an emergent symmetry that are amenable to progress with sufficient resources for lattice calculations.  One central issue concerns the nature of emergent symmetries generally.  Emergent symmetries involve the taking of limits.  The symmetry emerges only as some parameter approaches a limiting value.  For example in the spin-flavor symmetry in baryons, emerges only as the large $N_c$ limited is taken; for finite but large $N_c$ the symmetry is approximate\cite{GS1,GS2,DM1,DM2,DJM1,DJM2,TDC96}  It is not immediately clear how the limiting process associated with this supposed emergent symmetry works. Understanding this, may give important clues as to its origin.  In the studies reported to date, the pattern seems to be clear by the time the 10 lowest lying modes are removed from propagators and to have become very good by the time 30 modes are removed\cite{DGL2}.  Thus, one might be tempted to conclude  that the symmetry emerges in the limit of a large number of modes removed and is approximate when the number is large but finite.  

 However, it is not immediately clear that this is the right way to view things.  Note that lattice calculations by their nature involve other limits as well---in particular the infinite volume limit and the continuum limits.  It is possible  that the infinite volume limit is playing a critical role here.  The number of modes that need to be removed  could depend on the size of the system.  Removing one mode from a system that is a kilometer on each side could well have a much smaller effect in terms of establishing an emergent symmetry than removing one mode from a system that is a femtometer  on each side.  Thus, rather than depending on the number of modes, perhaps one should focus on the virtuality of the modes excluded--where the virtuality $\lambda$ is the eigenvalue of the Dirac operator.  Thus, it may seem plausible that condition  would be that the symmetry emerges in the limit where the virtuality of the excluded is large on some fixed scale as the volume is taken to be large.    The calculations to date have shown that correlation functions become similar  for a virtuality of about 65 MeV and look to be nearly identical given the quality of the numerics by 180 MeV.  
 
One interesting possibility is that in the limit that the size of the system goes to infinity, the effect might require a diverging number of modes to be removed while at the same time setting in at  decreasingly small virtuality.   Recall that, for large volume with many modes contributing, the notion of a density of modes is useful---in this regime $\frac{1}{V} \sum_n$ (where $n$ labels the mode)  can be replaced  by $\int d \lambda \rho(\lambda)$.  The condition for the emergent symmetry could be that enough modes are removed so that the full region associated with density of modes at zero is removed. This could require an increasing number of modes as $V$ gets large.  This is a condition worth considering in that the density at zero is associated with spontaneous chiral symmetry breaking due to the Banks-Casher relation.   If this turns out to be the appropriate condition, then the maximum virtuality of the excluded states need to produce the onset of an emergent symmetry will drop with increasing volume, since the key thing is the density at zero.  In any case,  with sufficient lattice studies at different volumes, one ought to be able to pin down the conditions that control the regime of validity for the emergent symmetry which could help elucidate its nature. 

Another issue that future lattice studies  should be able to resolve concerns the role of  chiral symmetry.  Recall that the originally studies in which quasi-zero modes were  removed were motivated by chiral symmetry and the fact that for zero quark mass,  the chiral condensate, an order parameter for spontaneous chiral symmetry breaking  is proportional to the density of Dirac modes near zero.  However, the remarkable fact at the heart of this paper is that correlation functions---when calculated with low-lying modes removed---in channels that are not connected by chiral symmetry are never-the-less identical.  This raises an obvious question which may go to the heart of understanding the pattern of  which  correlation functions will become identical: what role, if any, is played by chiral symmetry?   One possibility that needs to be explored is that chiral symmetry may be unconnected to the observed phenomena.  One reason to suspect this, is  that the symmetry pattern is  larger than given by chiral symmetry.   Another hint is the phenomenon is clearly observed even though the actual lattice calculations were done reasonably far from the chiral limit.  The pion mass  for the studies in refs.~\cite{DGL2,DGL1,DGP1,DGP2}  was more than twice the physical value.   A final reason to explore this possibility are the results of ref.~{cond1,cond2} which suggest that the chiral condensate remains nonzero even after the removal of $\sim 10-30$ low lying modes---a regime in which the emergent symmetry---appears to be manifest.  This is consistent with the possibility that the observed pattern of nearly identical correlators may be unconnected with chiral symmetry.

Fortunately, lattice calculations should be able to resolve the role of chiral symmetry provided that high quality calculations with differing quark masses are be done. If chiral symmetry is not necessary, the phenomenon should remain observable with similar quality for calculations done with large quark masses.    In contrast,  if chiral symmetry is necessary,  then  one would expect that the correlation functions would become increasingly similar as the quark mass is made smaller. 

There is one very interesting, but highly nontrivial type  of lattice calculation, that if done  might shed  light onto the issue.  It is worth recalling that the lattice calculations were done with the quasi-zero modes removed from propagators connected to sources, but were not removed in the calculations of the fermion determinant.   That is valence and sea quarks are treated on a different footing.   It would certainly be of interest to know whether the symmetry patterns would be the same if the gauge configurations were computed with fermion determinants that also had the zero modes excised.\\

\begin{acknowledgments}
The author thanks Leonid Glozman for introducing this issue to him and numerous insightful comments.  This work was begun at the University of Graz; their hospitality is great appreciated. The support of the US Department of Energy is also gratefully acknowledged. 
\end{acknowledgments}

\end{document}